\documentclass[twoside,journal]{IEEEtran}
\usepackage{amsmath,amssymb,amsfonts}
\usepackage{algorithmic}
\usepackage{algorithm}
\usepackage{array}
\usepackage{subcaption}
\usepackage{textcomp}
\usepackage{stfloats}
\usepackage{url}
\usepackage{verbatim}
\usepackage{graphicx}
\usepackage{cite}

\begin{document}

\title{Energy Efficiency Support for Software Defined Networks: a Serverless Computing Approach}




\author{
 \IEEEauthorblockN{Fatemeh Banaie\IEEEauthorrefmark{1}, Karim Djemame\IEEEauthorrefmark{2}, Abdulaziz Alhindi\IEEEauthorrefmark{2}\IEEEauthorrefmark{3}, Vasilios Kelefouras\IEEEauthorrefmark{4}} \\
 \IEEEauthorblockA{\IEEEauthorrefmark{1}School of Computing and Engineering, University of Huddersfield, UK} \\
\IEEEauthorblockA{\IEEEauthorrefmark{2}School of Computing, University of Leeds, UK} \\
\IEEEauthorblockA{\IEEEauthorrefmark{3}Department of Computer Science, Qassim University, Kingdom of Saudi Arabia} \\
\IEEEauthorblockA{\IEEEauthorrefmark{4}School of Engineering, Computing and Mathematics, University of Plymouth, UK}

\thanks{ This work was supported by the European Next Generation Internet Program for Open INTErnet Renovation (NGI-Pointer 2) under contract 871528 (EDGENESS Project).}
\thanks{K. Djemame is with the School of Computing, University of Leeds, Leeds LS2 9JT, UK (e-mail: K.Djemame@leeds.ac.uk). F. Banaie Heravan is with the School of Computing, University of Leeds, UK (e-mail: F. banaieheravan@leeds.ac.uk). A. Alhindi is with the Department of Computer Science, College of Computer, Qassim University, Kingdom of Saudi Arabia, and with the School of Computing, University of Leeds, UK, (e-mail: a.allhndi@qu.edu.sa). Vasilios Kelefouras is with the School of Engineering, Computing and Mathematics, University of Plymouth, UK (e-mail: Vasilios.kelefouras@plymouth.ac.uk).
}
}


\markboth{Submitted to IEEE Transactions on Services Computing,~Vol.~XX, No.~X, February ~2024}%
{Shell \MakeLowercase{\textit{et al.}}: A Sample Article Using IEEEtran.cls for IEEE Journals}

\IEEEpubid{0000--0000/00\$00.00~\copyright~2021 IEEE}

\maketitle


\begin{abstract}
Automatic network management strategies have become paramount for meeting the needs of innovative real-time and data-intensive applications, such as in the Internet of Things. However, meeting the ever-growing and fluctuating demands for data and services in such applications requires more than ever an efficient and scalable network resource management approach. Such approach should enable the automated provisioning of services while incentivising energy-efficient resource usage that expands throughout the edge-to-cloud continuum. This paper is the first to realise the concept of modular Software-Defined Networks based on serverless functions in an energy-aware environment. By adopting Function as a Service, the approach enables on-demand deployment of network functions, resulting in cost reduction through fine resource provisioning granularity. An analytical model is presented to approximate the service delivery time and power consumption, as well as an open source prototype implementation supported by an extensive experimental evaluation. The experiments demonstrate not only the practical applicability of the proposed approach but significant improvement in terms of energy efficiency.
\end{abstract}

\begin{IEEEkeywords}
Network management, Software Defined Networks, serverless computing, Network Function Virtualisation, energy efficiency.
\end{IEEEkeywords}

\section{Introduction}
\label{sec:introduction}
\subsection{Background}
\IEEEPARstart{T}{he} unprecedented growth of data traffic caused by the ever-expanding use of smart devices has highlighted the challenges of managing and controlling the information in existing networks \cite{Makh2022, Ban2018}. Indeed, the traditional rigid and static networking infrastructure was initially designed for a particular type of traffic, namely monotonous text-based content, which is not well suited to the ever-increasing interactive and dynamic multimedia streams. Additionally, the  emergence of the Internet of Things (IoT), with the provision of cloud computing tools and services, offers a new generation of advanced services with more stringent communication requirements for supporting innovative use cases. To meet the challenges of the explosive traffic demand, automatic and adaptive network management strategies can be applied to overcome the challenges in terms of scalability, reliability, and cost-effectiveness, thereby guaranteeing the user’s Quality of Experience (QoE) \cite{Bal2021, McC2019}.

Software Defined Networking (SDN) and Network Function Virtualisation (NFV) are the two promising technologies in realising the transition to next-generation IoT networks. SDN offers efficient resource management by separating the control logic from the hardware. Thus, it builds a flexible network management scheme, enabling a remote software controller to program the forwarding states in the data plane owing to a set of network policies. On the other hand, NFV provides a cost-efficient implementation of network functions by enabling shared physical resources either locally or on the remote cloud \cite{Bek2018}. The network virtualisation technique not only offers a reduced cost by sharing the network infrastructure but also improves time to market for novel applications. Therefore, these key tenets are critical for facilitating innovative applications, enabling network slicing, and ensuring a superior user experience (QoE) through simplified network management. 

Existing work has investigated the challenges in deploying SDN and the key factors for efficient service delivery to the end users \cite{Fal2016, Ust2020}. These studies cover various aspects such as the number of Virtual Network Functions (VNFs) running on a Virtual machine (VM), their resource demands, NFV-based service development, and the requirements in existing Service Development Kits (SDKs). Several open-source implementations of SDN, such as open network operating system (ONOS) \cite{Ons2014} and OpenDaylight (ODL) \cite{Med2014}, integrate the use of VNFs in service provisioning. These platforms adopt a monolithic software approach that utilizes NFV to implement agile service delivery on cloud infrastructures. However, the monolithic design of current SDN controllers aggregates it into a single  application that requires a specific set of programming interfaces. Hence, the development of new applications/services is restricted by making them dependent on a particular SDN controller and programming language. The adoption of microservices-based architectures has led the way towards a distributed SDN architecture that is designed to split the current monolithic controller software into a set of cooperating microservices \cite{Cor2019}. 

\subsection{Motivation}
A microservice-based SDN/NFV architecture plays a significant role in simplifying application development by eliminating specific interface requirements. Standard APIs connect users to the gateway, enabling on-demand services. Service functions (i.e., microservices) are deployed as VNF in a container in a distributed environment. Such container-based SDN platform enables the orchestration of tasks such as automation of service delivery, scaling up/down, and fault isolation. This enhances network flexibility and ensures quick adaptability to new services. Such services usually involve simple processing (e.g., packet processing/handling), which are \textit{short-lived} and \textit{event-driven}. 

\begin{figure}
\center
\includegraphics[width=7.7cm]{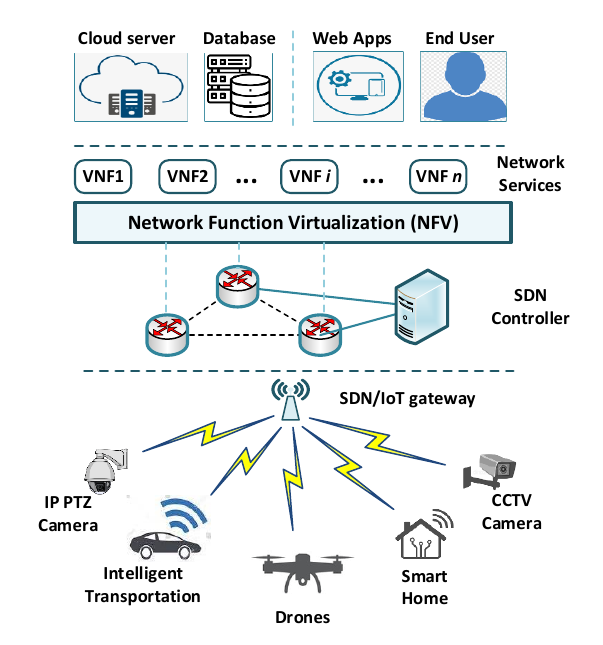}
\caption{Microservice-based IIoT service platform.}\label{fig0}
\end{figure}

Clearly, this service platform requires ubiquitous virtual resources with a dynamic and flexible service environment, with VNFs used to provide a service-specific overlay on top of the existing network topology, as shown in Figure \ref{fig0}. Thus, they have the potential of being executed on a serverless platform \cite{Li:2023}. Serverless computing, with the support of {\it Function as a Service (FaaS)}, is an ideal solution to build and optimise network services with zero infrastructure and maintenance costs as all of a server's typical operational back-end responsibilities are offloaded. It also decreases operational costs by offering a finer granularity of resource provisioning that is a driving factor for energy efficiency.

As the serverless paradigm provides a resource-efficient alternative to virtual machines (VMs) and containers, it can therefore effectively support SDNs and function virtualisation. Notably, serverless computing is well-suited to SDN controllers as they are highly event-driven, modular, and parallel \cite{Adi2019}. 

The innovation lies in the incorporation of serverless architectures with 1) SDN controllers, with minimal sharing of state between the modules, and 2) NFV for orchestrating VNF as well as applications that require running short, on-demand tasks operating on data collected from the data plane. VNFs launched and orchestrated in a serverless manner incentivises efficient resource usage and provides granular reporting on a function level: functions take up the most execution time can be identified as well the associated {\it cost} for running them. This is essentially a proxy for energy usage as a unit of (serverless) compute, making VNFs instantiation and orchestration significantly energy and resource efficient \cite{Djemame:2021}.




\subsection{Contributions}

This paper introduces a novel approach that leverages the serverless platform to develop a modular and microservice-based SDN architecture, offering energy-efficient and scalable service provisioning in the network. 
To the best of our knowledge, this is the first work aiming at realising the concept of modular SDN based on serverless functions in an energy-aware environment. In our previous work \cite{Banaie:2022}, the initial results are presented as proof of concept for SDN/NFV-based service delivery in the network. 
 This paper, however, goes further by presenting the proposed architecture through analytical and experimental modelling. The main contributions of this paper are outlined as follows:
\begin{itemize}
    \item 
    a microservice-based network management approach using SDN and NFV technologies is proposed. The method divides network services to ensure that the SDN controller delivers only essential functionalities (e.g., flow and topology services). Other network services, such as firewall, traffic monitoring, and IoT services, can be deployed on the serverless platform, with a little-to-no operating expense;
    
    \item a modular and distributed SDN architecture in which the distribution of events is supported by external processes is developed. This method leverages the benefits of a serverless platform to provide energy-efficient and scalable services for next-generation innovative applications, paving the way for more efficient and sustainable technological advancements;
    
    \item an analytical model is presented to approximate the service delivery time and power consumption;  

    \item a prototype implementation of the proposed architecture is developed and supported by a real-world experimental setting that compares the performance of modular SDN with a Docker containers \cite{docker2} network deployment. The evaluation results demonstrate supporting measurements such as power efficiency and latency;
    
    \item the open-source platform of modular SDN, using well-known open-source software such as ONOS \cite{Ons2014}, OpenFaaS \cite{Opn2021}, and gRPC \cite{Grp2016}, is implemented and available on GitHub for the research community\footnote{https://github.com/EDGNSS}.
    
\end{itemize}

The rest of the paper is organised as follows. Section 2 provides an  overview of the related literature. Section 3 describes the proposed architecture along with the functional definition of its components and interfaces. In section 4 we present the analytical model to approximate the service delivery time and power consumption, and this is followed by its evaluation in section 5. The modular SDN prototype implementation is explained in section 6, and performance results discussed in section 7. Section 8 concludes the paper. 

\section{Related Work}
Improving the energy efficiency of SDNs has been a research issue since SDN was introduced to the market. Rout \cite{rltd1} gives a wide overview of the popular methods of improving energy efficiency in SDN. These methods can be classified into three categories traffic aware, end-host aware and rule placement. The main point in the first and the second category is to monitor the network traffic in order to select the devices and links that could be turned off without affecting performance, which, in turn, would decrease the energy consumption of SDNs. The third category, however, focuses on how to save energy by controlling the number of flow entries in Ternary Content Addressable Memory (TCAM). The main part of SDN is the controller, and controllers are usually developed as monolithic programs that preform all the tasks of the control plane. Therefore, there have been a number of research papers discussing the idea of disaggregating the controller into separate units, which, in turn, would ease maintaining and developing the controller as well as give a potential of improving its energy consumption. 

Maity \cite{Maity:2023} addresses one of the aspects of the control placement problem that
usually includes dividing the network into a number of sub-networks, depending on
the size of the network, and assigning the management of partition to a single
controller. The proposed approach aims to find the optimal number of sub-networks
considering that control messages are exchanged through the in-band control plane
and to deactivate unused links as well as reroute flows through energy-aware
approaches. To find the optimal number of controllers, a mathematical model based on a coalition game generates the number of sub-networks considering the trade-off between switch-to-controller latency and
energy consumption. The evaluation of the proposed solution shows that it is capable of saving energy by roughly 22\%. 

Oliveira \cite{Oliveira:2021} discuss how to improve the energy efficiency of SDN
controllers by making use of multi-core processors running at a lower
frequency with the aim to consume less power than with a single core running at a higher
frequency. An end-host-aware solution exploiting a parallel implementation of the SDN controller to distribute its load between multiple processor cores
while lowering their frequency to trade-off performance against energy consumption is shown to achieve 28\% drop in energy consumption.

The work in \cite{Priyadarsini:2020} discusses another aspect of SDN’s energy efficiency that
is related to an energy-efficient method for balancing the incoming loads between
controllers. Their traffic-aware solution is designed to migrate some of the loads on
highly loaded controllers to lightly loaded ones, which includes migrating some of the
switches as well. The procedure of migrating switches requires an energy-efficient
re-routing algorithm for reducing energy consumption while ensuring the service
level is not violated. They design a system consisting of a number of components
that measure the load of each controller continuously and ensure that each controller
is aware of others’ loads. Therefore, each controller can calculate and update the
threshold at which the component of migrating loads is triggered. Migrating switches
and their load can happen only through energy-efficient paths that are chosen before
the migration procedure begins, which needs to avoid affecting the service level of
the network. The energy-efficient paths are selected by their proposed model,
EERAS, whose task is to select the shortest path with minimum energy consumption
between the two controllers involved in the migration process. The evaluation of their
experiment shows that the proposed solution achieved roughly a 25\% reduction of
energy consumption as well as a 20\% improvement in performance.

Djemame \cite{Djemame:2021} discusses that the deployment of Virtualised Network Functions (VNFs) on a serverless platform could lead to an improvement in the energy consumption of SDN/NFV architecture. As serverless functions are event-driven and running short, they are likely to be more efficient in resource usage as well energy consumption. The paper argues that an energy-aware placement of VNFs could reduce energy consumption as well as the latency, such as VNFs belonging to the session might be deployed in the same server to reduce the time of transferring data. 

Banaie and Djemame \cite{Banaie:2022} describe how the integration of serverless with a SDN/NFV architecture could decrease the latency when the SDN controller is divided into separate modules deployed and executed as serverless functions. This suggested solution leverages the technology of virtualisation to allow to deploy network functions as serverless functions. The SDN controller core is kept to provide only the minimum required functionality, while the other services exist outside of of the controller where deployed on a serverless platform. 


There have been a multitude of research investigations in the area of energy consumption in the serverless space. 
Alhindi et al. shows in \cite{Alhindi:2022} that a serverless platform (OpenFaaS) is more efficient in consuming power compared to Docker containers. This is demonstrated through a number of experiments conducted to both platforms when they receive intensive loads that are designed to stress the CPU and memory resources. Moreover, 
OpenFaaS with faasd tends to consume less power on a memory-heavy benchmark than its Kubernetes \cite{kube2} counterpart with a decrease of 58\%, which is expected, as the runtime is more lightweight. 

Cicconetti et al. \cite{Cicconetti:2021}, proposes a framework for efficient dispatching of stateless tasks with the goal of minimising the response times and exhibiting short- and long-term fairness. Their evaluation on OpenWisk platform shows that the interaction with SDN controller can be useful relieving the network congestion. A high-performance serverless platform for NFV is presented in \cite{Shen:2020}, in which the authors utilize three different mechanisms for minimising the latency, including state management, efficient NF execution model, and avoiding packet latency. The work presented in \cite{Tzenetopoulos:2021} proposes a methodology for application decomposition into fine-grained functions and energy-aware function placement on a cluster of edge computing devices subject to user-specified QoS guarantees.

Jia and Zhao \cite{Jia:2021} propose a mechanism for energy-aware resource allocation in a serverless context to minimise power consumption named RAEF. An agent running at the function level is proposed, constructed of four components, where the most important are the predictor and the resource explorer. The findings show that the proposed solution can reduce energy consumption anywhere from a noticeable 9.7\% to a significant 21.2\% across different workloads. Moreover, the control-plane components of OpenFaaS consumes an insignificant amount of power compared to the function runtime plane.

A software-hardware co-design solution named DAC (Differentiate and Consolidate) that is designed to schedule serverless functions based on their invocation interval time(IIT) is proposed by Zheng \cite{Zhang:2022}. The solution aims to improve the energy
efficiency of the serverless framework by placing the serverless functions according to their categories: native (power and non-power sensitive) and tail functions (likely to suffer from cold starts). The consolidation controller uses an adaptive algorithm that considers two parameters, {\it B-warm} which refers to the benefits of functions’ warm-start and {\it P-warm} which is the warm-start probability. The solution is shown to reduce the energy consumption of tail functions by up to 23\%.

The work of Das et al. \cite{Das:2020} focuses on cost efficient execution of multi-function serverless applications on hybrid cloud deployments based on OpenFaaS and AWS Lambda. A framework named Skedulix is proposed based on a greedy solution to scheduling after it was modelled into a Mixed Integer Linear Program (MILP). Serverless applications are modelled as directed acyclic graphs (DAGs), in which each node represents a different function. The goal of the scheduler is to speed up processing at the lowest cost possible, by maximising the use of the private cluster and offloading any incompletable work to the public cloud in cases where the deadline cannot be met. 

Fan and He \cite{Fan:2020} target the well-known issue of cold start by devising a new scheduling strategy that allows creating multiple pods at a time instead of the default pod-by-pod approach that Kubernetes adopts as it has to traverse each node to compute the score that determines the most beneficial placement which can cause a slowness in start-up latency. Essentially, they propose a simple algorithm based on Mixed Integer Programming (MILP), that divides pods into groups and schedules a group at a time when it passes a set overload threshold and achieves sizeable reduction in latency in simulation, ranging from $\sim$20-60\% depending on the number of pods that need to be scheduled.

In summary, existing work has investigated the challenges in deploying SDNs in a serverless computing platform as well as the key factors for efficient service delivery. This paper goes further as it aims at realising the concept of modular SDN based on serverless functions in an energy-aware environment. This research is backed up by an analytical model and analysis of the architecture as well as its cloud-based prototype implementation and evaluation.

\section{Microservice-based Provisioning of Network Services}

This section presents the proposed microservice-based network management approach. The innovation lies in the utilisation of network programmability within the context of Network Function Virtualisation (NFV) and the exploration of the benefits the serverless computing paradigm offers \cite{Fal2016}.  As noted, SDN offers scalable network management by separating the control plane and data plane in a network. The control plane consists of two main components: (i) SDN core functionalities, such as flow services and topology services, and (ii) management applications, such as traffic monitoring, load-balancing, and firewall. These components communicate requirements via Northbound and Southbound Application Program Interfaces (APIs). 
However, the centralised management of the network raises potential issues in terms of failure, scalability, and service latency, which aims to be resolved by distributed deployment of the controller \cite{Ban2018, Cor2019, Arz2022}. Existing SDN controllers adopt a monolithic design that aggregates all functions into a single program. This approach constrains its ability to deploy a new service independently from other services. Unlike existing monolithic architectures, the proposed microservice-based SDN Architecture adopts a modular and flexible approach by breaking down SDN functionalities into independent microservices.
\begin{figure}
\hspace{-0.7cm}
\includegraphics[width=9.3cm]{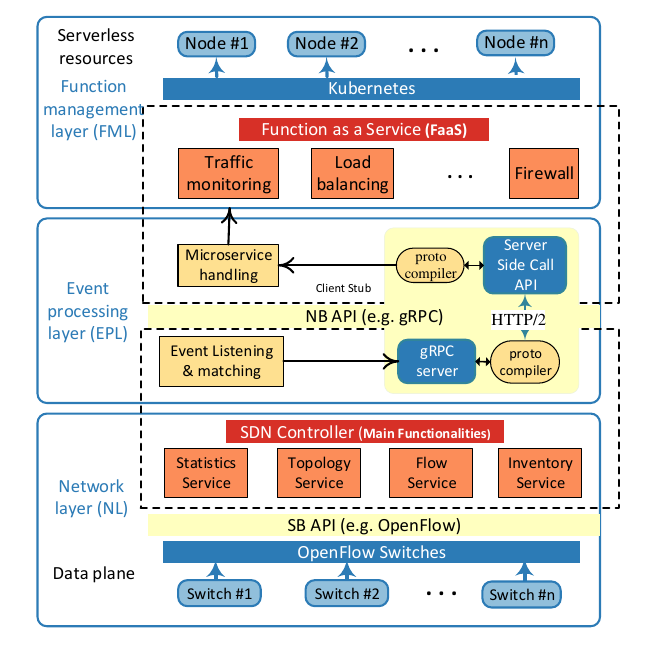}
\caption{SDN/NFV architecture in a serverless environment.}\label{fig1}
\end{figure}

For this objective, we have developed the controller as a composition of containerised services, tightly integrated and collaborating seamlessly. By adopting this strategy, the FaaS-enabled architecture promotes a modular and scalable design that enables efficient development and management of network services. Next, a functional definition of its components and interfaces is presented.

\subsection{Architecture}
The modular and distributed SDN architecture utilises disaggregated software modules deployed on a serverless platform as depicted in figure~\ref{fig1}. This model leverages the benefits of NFV to decouple networking software from the hardware that delivers it. Therefore, the software modules (i.e., management applications) can be deployed as independent microservices on the serverless platform. In this approach, SDN core services provide minimum required functionalities, and the other services can be developed as external applications in the form of collaborative software modules. Such distributed SDN/NFV architecture consists of three operational layers including the network layer (NL), event processing layer (EPL), and virtual functions management layer (VML). 

\subsection{ Network Layer} The network layer is responsible for establishing network rules and exchanging information among network elements. This layer consists of two components: (i) a data plane, i.e., OpenFlow-enabled switches, and (ii) a control plane supporting the SDN core functionalities such as topology service, flow service, inventory service, etc.

The service structure in modular SDN utilizes an \textit{event distribution} system to externalize event processing in SDN. This approach enables the division of control plane services into a set of cooperating microservices. In this method, the SDN controller operates by processing events received from both the southbound and northbound APIs. An event is defined as any changes in the network that invokes one or more management applications. Therefore, the SDN controller can be developed as an \textit{event-driven} and \textit{modular} software responding to these events. In Particular, the events play a crucial role in informing applications about network changes. This aim is realised by adding an \textit{event listener module} that listens to the events coming from the underlying network. Upon receiving a new event, the network will handle invoking the corresponding functions on the serverless side. Thus, the management applications deployed on the serverless platform can be executed on demand as needed.   

\subsection{Event Processing Layer} This layer enables the externalisation of event processing in SDN by providing efficient and reliable communication among different services or components within an application or across multiple applications. The Event processing layer (EPL) consists of three main components including:
\begin{itemize}
    \item [-] \textit{Event listening module} in the SDN layer is responsible for capturing OpenFlow-based events originating from the underlying network and forwarding them to the serverless connector.
    \item[-] An \textit{Event processing module} that is responsible for using a communication interface to notify SDN events to the Virtual Functions (VFs) deployed on a serverless platform. This module enables interaction with virtual functions via a communication interface/protocol.
    \item [-] A \textit{ Microservices handling module} in the serverless layer that manages functions and invocations by receiving event notifications from the underlying network
\end{itemize}

As the main principle is effective decomposition, where network information and state remain synchronised and consistent, this layer provides an effective interaction of the functions with each other or with the event-processing module of the controller (i.e., sending and receiving the event notifications among microservices). This method enables separate component development, promoting efficient reuse across diverse deployments. Thus, it facilitates modularity, collaboration, and scalability in SDN architectures.  


\subsection{Function Management Layer} The FaaS platform allows developers to deploy functions that are triggered by specific events or requests, including packets arriving at the network layer. The functions invoked by these events can process the packets and create respective responses for them. The advantage of using FaaS for packet processing is that developers can focus on writing the logic specific to their functions, without the need to manage low-level networking infrastructure or worry about scalability. The FaaS platform handles the scaling, resource allocation, and execution of the functions based on the incoming workload.

\section{Analytical Model} 

In Section III, we outlined the details of the microservice-based network management through the modular SDN architecture. In this section, we will discuss the details of the analytical model to approximate the service delivery time and power consumption, based on the system described earlier. Our primary focus here is to obtain the steady-state metrics of the proposed system based on the system configurations and workload characteristics.

\begin{figure}
\center
\includegraphics[width=7.48cm]{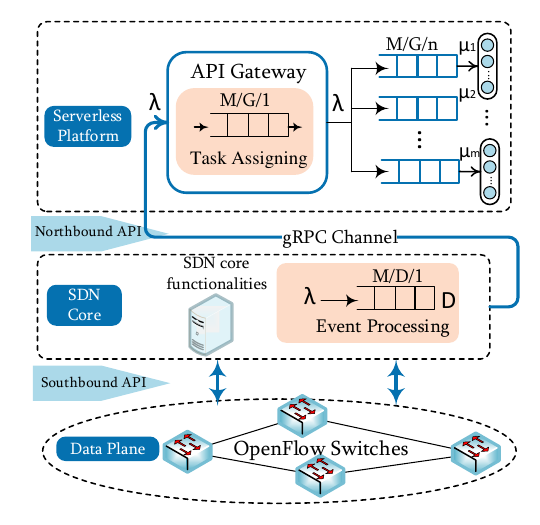}
\caption{System Model for distributed SDN.}\label{fig2}
\end{figure}

The performance model is illustrated in figure \ref{fig2}. Let us assume that the network operates in consecutive time slots $\{T_\tau|\tau\!=\!\!1,2,3,...\}$ in which a set of $m$ functions $ \mathcal{F}\!\!=\!\{f_1,f_2,...,f_m\}$ are subscribed for events in the gateway. Similar to many other works of literature \cite{Tan2020, Tan2022}, a quasi-static scenario is adopted where the environment is considered static at each time slot but it might vary in different time slots. 


After receiving a new event packet, the controller processes it and then forwards the packet toward the edge gateway. The edge gateway, in response, triggers the corresponding subscribed function on the edge computing nodes. Note that we will use packets and events interchangeably in the rest of the paper as the arriving packets from network elements are used to notify the events and changes in the network. 
The modularity of the system allows for deploying functions in a distributed environment and scaling them according to the service workload. This capability enables the initiation of multiple instances of the same service (i.e. multiple replicas of the functions), effectively addressing latency issues in microservices. 
We calculate steady-state estimates for various deployment characteristics based on the workload characteristics and the output model. Key notations for the most important parameters are summarised in Table \ref{t1}.

\subsection{Delay Model}

As the population size of a typical network/IoT application is relatively high, and while the probability of a given user requesting a service is relatively small, the arrival process can be modeled as a Markovian process \cite{Gri2010}. Therefore, we assume that the packet arrival rate for each function $m$ follows a Poisson Point Process with the average rate of $\lambda_m$, which indicates that the event inter-arrival time is exponentially distributed with a rate of $1/\lambda_m$. Accordingly, the aggregated arrival rate for all functions is also Poisson and can be denoted as
\begin{eqnarray}
\lambda\!\!=\!\sum_{m=1}^M \lambda_m,\quad \forall m \!\in\! \mathcal{F}
\end{eqnarray}

The \textit{event processing module} serves the arriving packets in non-preemptive First Come First Served (FCFS) discipline and directs them toward the edge gateway for more processing. The goal is to compute the latency of the service delivery on edge computing resources by receiving an event notification from the underlying network. To do so, we need to analyse the total latency of the computation and transmission entities of the network, and propose the analytical model to approximate the service delivery time and power consumption.

\subsubsection{Transmission Delay of Controller} 
As discussed, the event notification packets from underlying network elements can follow the Poison process with the aggregated arrival rate of $\lambda$ at time slot $\tau$. Meanwhile, denote $\mu_m$ as the transmission delay of the event processing module for the function $m$, which is deterministic and depends on the resource capacity of the controller. For simplicity, we assume the packet size $\phi_m$ is the same for all types of events within all functions, i.e., $\phi\!\!=\!\phi_m$, for each $m\!\in\!\mathcal{F}$.  As $\phi_m$ is constant for all functions, $\mu_m$ is also constant, i.e., $\mu\!\!=\!\!\mu_m$. Hence, the transmission delay can be calculated as $\mu\!=\frac{\phi}{\mathcal{R}}$, where $\mathcal{R}$ is the wireless transmission rate between controller and edge gateway and can be calculated as \cite{Rio2014}
\begin{eqnarray}
\mathcal{R}=\mathcal{B}log_2(1+\frac{\mathcal{G} w}{\sigma^2}),
\end{eqnarray} 
Here, $\mathcal{B}$ is the bandwidth, $\mathcal{G}$ is the wireless channel gain, $w$ is the transmitted signal power, and $\sigma$ is the noise power. Given this value, the event processing of the controller can be considered as a $M/D/1$ queueing model with a deterministic service time of $\mathcal{D}=1/\mu$ for all functions. Thus, the average service latency of the controller (i.e., the waiting time in the queue plus the transmission time) for each class $p$ can be derived as
\begin{eqnarray}
\overline{TS}(\tau)\!\!=\!\!\!\underbrace{\frac{\rho}{2\mu(1-\rho)}}_{waiting-time}+\!\!\!\!\!\!\!\!\!\!\underbrace{1/\mu}_{transmission-time}\!\!\!\!\!\!\!\!\!\!\!\!\!\!\!,
\end{eqnarray}
where $\rho$ is the offered load of the controller and can be derived as $\rho=(\sum_{m=1}^M \lambda_m)/\mu$.

\begin{table}[h]
\caption{Parameter notation}
\label{t1}       
\centering
\begin{tabular}{l|l}
\hline\hline
Parameter & Description  \\
\noalign{}\hline\noalign{}
\quad$\tau$ & Discrete time slots  \\
\quad$m$ & Number of Functions  \\
\quad$n$ & Instances of Functions  \\
\quad$\lambda$ & Total arrival rate of events \\
\quad$\mu$ & Mean service rate of events\\
\quad$\rho$ & Offered load of controller \\ 
\quad$\phi$& The packet size\\
\quad$\mathcal{R}$& Wireless transmission rate\\
\quad$\mathcal{B}$& Bandwidth available for channel\\
\quad$\mathcal{G}$& Wireless channel gain\\
\quad$\eta$&Power coefficient\\
\quad$x_i$&binary variable indicating cold/warm start\\
\quad$w$&transmission signal power\\
\quad$\sigma$& noise power\\
\quad$\overline{TS}$ & Mean transmission delay of the controller \\
\quad$\overline{Wg}$ & Mean waiting time in the queue for gateway\\
\quad$\overline{Tg}$ & Mean response time of the gateway\\
\quad$\overline{Wf}$ & Mean waiting time for function $m$\\
\quad$\overline{Tf}$ & Average response time of function $m$\\
\quad$\overline{T}_m$ & Mean response time of the network for function $m$\\
\quad$U_s$& CPU utilisation of server $s$\\
\quad$C_s$& Number of cores for server $s$\\
\quad$F_s$& CPU cycle frequency of server $s$\\
\quad$D_s$& Cold container startup delay of server $s$\\
\quad$Pidle$& Static power consumption of server $s$\\
\quad$Pdyn$& Dynamic power consumption of server $s$\\
\quad$Pw$& Power consumption for warm container startup\\
\quad$Pc$& Power consumption for cold container startup\\
\quad$P_{s}$& Total power consumption of server $s$ \\
\hline\hline
\end{tabular}
\end{table}

\subsubsection{Processing Delay of Gateway}
Algorithm 1 shows the pseudo-code of service provisioning on the serverless platform, which can be divided into three consecutive intervals for each time slot $\tau$. Once an event is received from the controller, the gateway queries the list of functions and their interesting events, \textit{function mapping interval}. It then maps the events into the respective function instances in the \textit{task assigning interval}, which leads to rendering the corresponding service by invoking the functions in \textit{execution interval}. 

The edge gateway plays an important role in the service delivery process as it connects the core of the network, i.e., the SDN controller, to the external applications deployed as microservices on the serverless edge platform. This module consists of two sub-modules: a \textit{function mapping module} which contains a list of functions and the events that they subscribed to them, and a \textit{task assigning module} for assigning the events to the function instances and deciding about the scaling up/down of the instances according to the workload. 

Hence, the processing time of the gateway can be modeled as an $M/G/1$ queuing system with the mean arrival rate of $\lambda$ and independent and identically distributed (i.i.d) general service time with mean $\overline{b}$ at time $\tau$. Thus, the total load is $\sum_{m=1}^M\rho_m$. The average waiting time of an event in the queue is \cite{Ros2014}:
\begin{eqnarray}
\overline{Wg}(\tau)\!\!=\! \frac{\lambda b^{(2)}}{2(1-\rho)}.
\end{eqnarray}

Here, $b^{(2)}\!\!=\!\! -\frac{d^2}{ds^2}{Bg}^*(0)$, where ${Bg}^*$ is the Laplace Stieltjes transform (LST) of the service time for each of the functions. Consequently, the average delay for service delivery can be obtained as $\overline{Tg}\!\!=\overline{Wg} + \overline{Bg}$ where $\overline{Wg}$ is the mean waiting time in the queues and $\overline{Bg}\!\!=\!-\frac{d}{ds}{Bg}^*(0)\!\!=\!b$ is the average service time of edge gateway, which is denoted by
\begin{eqnarray}
\overline{Tg}(\tau)\!\!=\! \frac{\lambda b^{(2)}}{2(1-\rho)} \!\!+\! b.
\end{eqnarray}

The average event processing time $\overline{Tp}$ of the functions at time slot $\tau$ can then be obtained by summing up $\overline{Ts}$ and $\overline{Tg}$, i.e.,
\begin{eqnarray}
\overline{Tp}(\tau)\!\!=\!\overline{Ts}(\tau)+\overline{Tg}(\tau), \quad \forall \tau \!\!\in\!T_{\tau}
\end{eqnarray}

\begin{algorithm}
 \caption{Service provisioning on FaaS}\label{A1}
 \begin{algorithmic}[1]
 \renewcommand{\algorithmicrequire}{\textbf{Input:}}
 \renewcommand{\algorithmicensure}{\textbf{Output:}}
 \REQUIRE  Functions $\mathcal{F}=\{f_m\}$, events $e$ 
 \ENSURE  $T_m(\tau)$ 
 \\ \textit{Initialisation} :
  \STATE Set values of parameters: $\lambda(\tau), \mu, m, \mathcal{I}$
  \STATE Initialize  $T_m(\tau), \tau$ and set $n_m=0$
  \FOR{ each event $e$} 
  \STATE Map $e$ to corresponding function $m$
  \IF{$n_m >0 $}
  \STATE Add event to the function queue
  \STATE calculate $W_m(\tau, n_m)$ with Eq.(8)
  \IF {$W_m \geq$ t} 
  \STATE $n_m \leftarrow n_m +1$ \COMMENT{Adding new replica for scaling up, t is a threshold for delay}
  \STATE Update $W_m(\tau, n_m)$ and $\overline{T}_m(\tau, n_m)$ with Eq.(8) and (10)
  \ELSE
  \STATE calculate $\overline{T}_m(\tau, n_m)$ with Eq.(10)
  \ENDIF
  \STATE Update parameters $\mathcal{I}=\{n_1, ..n_m\}$
  \ENDIF  
  \ENDFOR
  \STATE Update parameters $\tau, \lambda$
  \RETURN $\overline{T}_m(\tau, n_m)$
 \end{algorithmic} 
 \end{algorithm}
 
\subsubsection{Processing Delay of Serverless Nodes}
The event packets served by the gateway can trigger the services, which may require invoking multiple replicas of the same function or instantiating a new instance of additional services. Let $n\!\!\in\!\mathcal{I}\!\!=\!\!\{n_1,n_2,...,n_m\}$ denotes the number of instances for the functions independently rendering network services to the event packets. In this case, each function can be modeled as an $M/G/n$ queueing system in which the events arrive at Poisson rate $\lambda_m$ and are served by any of $n$ instances, each of whom has the service distribution $G$.  Note that $n\!\!=\!0$ indicates that there is not any running instance of the function $m$ in the system. As approximated in \cite{Ros2014}, we have 
\begin{eqnarray}
&&\overline {Wf}_m(\tau, n)\!\!=  \\
&&\!\frac{\lambda_m^{n} \, {b_m}^{(2)}(b_m)^{n-1}}{2(n\!-\!1)!(n\!-\!\lambda_m b_m)^2 [\sum_{i=0}^{n-1} \frac{(\lambda_m b_m)^i}{i!}\!\!+\!\frac{(\lambda_m b_m)^n}{(n-1)!(n-\lambda_m b_m)}]}, \nonumber
\end{eqnarray}
where $n=n_m$ for function $m$, $b_m$ and ${b_m}^{(2)}$ are calculated as $b_m\!\!=\!\!-\frac{d}{ds}{B_m}^*\!\!(0)$ and  $b^{(2)}_m\!\!=\!\! -\frac{d^2}{ds^2}{B_m}^*\!\!(0)$, respectively. Let us assume that the service time of the events follows the exponential distribution with mean $\mu_m$ for each function $m$. Then, we have
\begin{eqnarray}
&&\overline{Wf}_m(\tau, n)\!\!= \\
&&\!\frac{(\rho_m)^n}{2(n-1)!\,\mu_m \,(n-\rho_m)^2[\sum_{i=0}^{n-1} \frac{(\rho_m)^i}{i!}+\frac{(\rho_m)^n}{(n-1)!(n-\rho_m)}]},\nonumber
\end{eqnarray}

Note that $\rho_m\!\!=\!\!\frac{\lambda_m}{\mu_m}$ is the offered load of instances for each function. This indicates that the utilisation of each function can be calculated as the ratio of the mean number of busy instances to $n_m$, which is obtained by $U_m\!\!=\!\!\rho_m/n_m\!\!=\!\!\lambda_m/n_m\mu_m$. Hence, the total latency of service delivery for the instances of the function $m$ at a serverless platform can be obtained by
\begin{eqnarray}
\overline{Tf}_m(\tau, n)\!\!=\!\overline{Wf}(\tau, n)+b, \quad \forall n\!\!\in\! \mathcal{I}, \tau \!\!\in\! T_{\tau}
\end{eqnarray}

Accordingly, we can obtain the total latency of packet processing for the function $m$ in this architecture as
\begin{eqnarray}
\overline{T}_m(\tau, n)\!\!=\!\overline{Ts}(\tau)\!+\!\overline{Tg}(\tau)\!+\!\overline{Tf}_m(\tau, n)
\end{eqnarray}

\begin{figure*}
     \centering
     \begin{subfigure}[b]{0.65\textwidth}
     \hspace*{+2cm}
     \center
         \includegraphics[width=9cm]{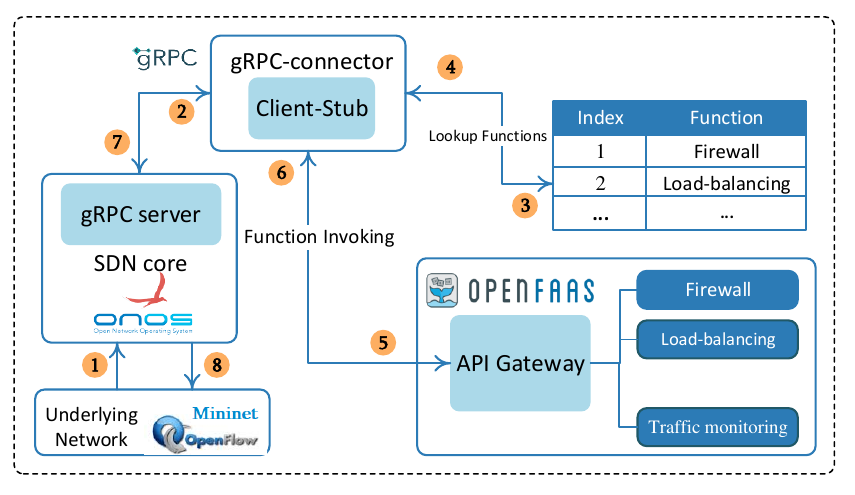}
         \caption{}
         \label{fig3}
     \end{subfigure}
     \begin{subfigure}[b]{0.3\textwidth}
         \includegraphics[width=4.5cm]{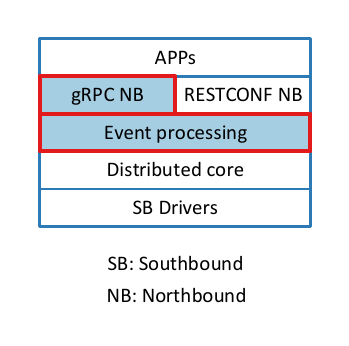}
         \caption{}
         \label{fig4}
     \end{subfigure}
     \caption{Implementing the microservice-based network architecture (a) ONOS and OpenFaaS integration sequence diagram. (b) Protocol stack of ONOS.}
\end{figure*}

\subsection{Power Model}
 The power consumption model for the function execution in a serverless edge computing environment can be considered in two different scenarios: a \textit{warm container} is ready to execute the function, and a \textit{cold start} in which a new container must be started prior to the execution \cite{mohan:2019}. Namely, \textit{warm start} refers to reusing an existing function instance, while \textit{cold start} involves launching a new function instance. During a cold start, the platform performs activities like initialising a new container, configuring the runtime environment, and deploying the function, which takes more time to process requests compared to warm starts. In the former case, the average energy expenditure of a function with a warm start is related to the task type and the power consumption of different sub-components of the server, e.g. the amount of processor, memory, and disk usage for different types of tasks \cite{Zho2022, Zha2018}. The power consumption of a running server $s$ (i.e., physical machine (PM)) can be divided into two components: the static part and the dynamic part. $Pidle$ represents a constant value that characterises the static part when the PM is not executing any tasks. In other words, if the PM is powered on and its utilisation is nearly zero, the power consumption will be equal to $Pidle$. The dynamic power consumption can be described by a quadratic polynomial model \cite{Aldossary:2019}. Thus, the power consumption is approximated by:
\begin{eqnarray}
Pdyn_s(\tau)=\gamma_1 U_s(\tau)+\gamma_2 U_s(\tau)^2, \quad \forall \tau\!\in\! T_{\tau}
\end{eqnarray}
where $U$ is the CPU utilisation for the running server at time slot $\tau$, and $\gamma_1$ and $\gamma_2$ are model parameters, the values of which are determined by linear regression during model generation. The power consumption of a running server can be modeled as
\begin{eqnarray}
Pw_s(\tau)=Pidle(\tau)+Pdyn(\tau), \quad \forall \tau\!\in\! T_{\tau}
\end{eqnarray}

For the latter one, the cold container startup time should be considered along with the power for function execution. Then, the power consumption for cold starting containers at time slot $\tau$ is

\begin{eqnarray}
Pc_s(\tau)=\eta F_s(\tau)^3 C_s(\tau) D_s, \quad \forall \tau \in T_{\tau}
\end{eqnarray}

Here, $\eta$ is the power coefficient, $F$ is the CPU cycle frequency at time slot $\tau$, $C$ is the number of containers that require a cold start at time slot $\tau$, and $D_s$ denotes the cold start delay for launching an instance $n$ on one of the server's cores \cite{Zha2018}. Thus, the power consumption of the server at time slot $\tau$ can be expressed as:
\begin{eqnarray}
P_s(\tau)=Pw_s(\tau)+\sum_{i=1}^k x_i(\tau) Pc_s(\tau),\quad \forall \tau \in T_{\tau}
\end{eqnarray}
where $k$ is the maximum number of running containers on the servers, $x_{i}\!\!\in\!\! \{0,1\}$ is a binary variable indicating whether the the $i^{th}$ container requires cold start at time slot $\tau$ (i.e., $ x_{i}\!\!=\!1$) or not (i.e., $x_{i}\!\!=\!0$).

\section{Model Evaluation}

In this section, we present an evaluation of the analytical model to offer deep insights into its underlying mathematical principles and properties as well as behavior and performance. The analytical model presented above has been solved using Maple 16 from Maplesoft, Inc., \cite{Map2015}. Similar to \cite{Ste2022, Mah2022, Sri2018}, we assume that event packets arrive in the system according to the Poisson process with the rate $\lambda$ and an exponential distribution of service time with mean $\mu$. The parameter $\lambda$ is varying from 20 to 100 arrivals per second and all function instances are assigned an identical service time duration, which is set to $(1/\mu)\!\!=\! 200ms$. The cold start delay, representing the time taken to launch a container on serverless edge computing nodes, is established within the range of $[0.15, 0.85]$ seconds. Furthermore, the computational capacity of individual containers is randomly distributed, spanning from $1 Gcycle/s$ to $2 Gcycle/s$, with an average of $1.5 Gcycle/s$.  
To assess and verify our performance models, two scenarios are considered. 

\begin{figure*}
     \centering
     \begin{subfigure}[b]{0.3\textwidth}
         \includegraphics[width=5.4cm]{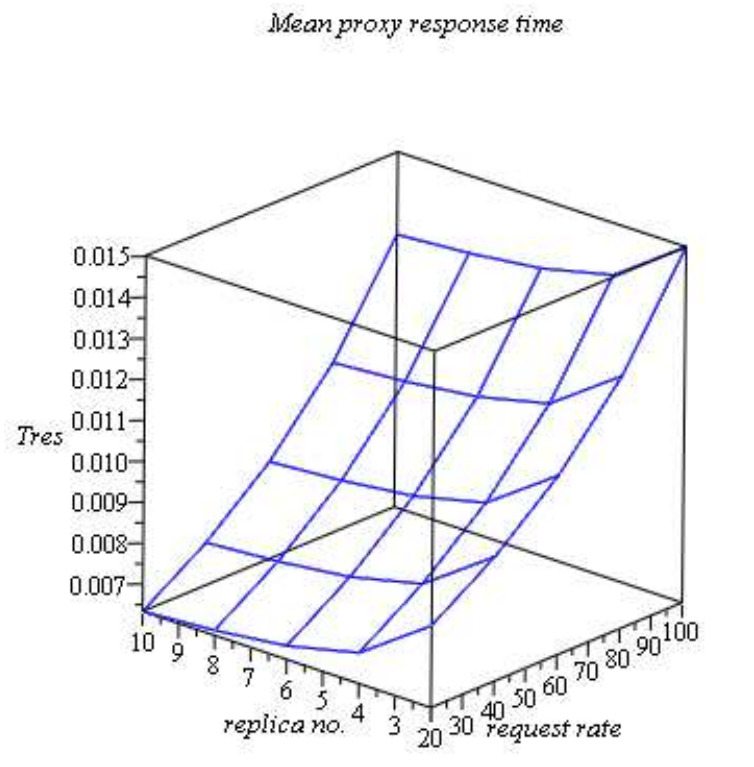}
         \caption{}
         \label{fig8a}
     \end{subfigure}
     \begin{subfigure}[b]{0.3\textwidth}
         \includegraphics[width=5.4cm]{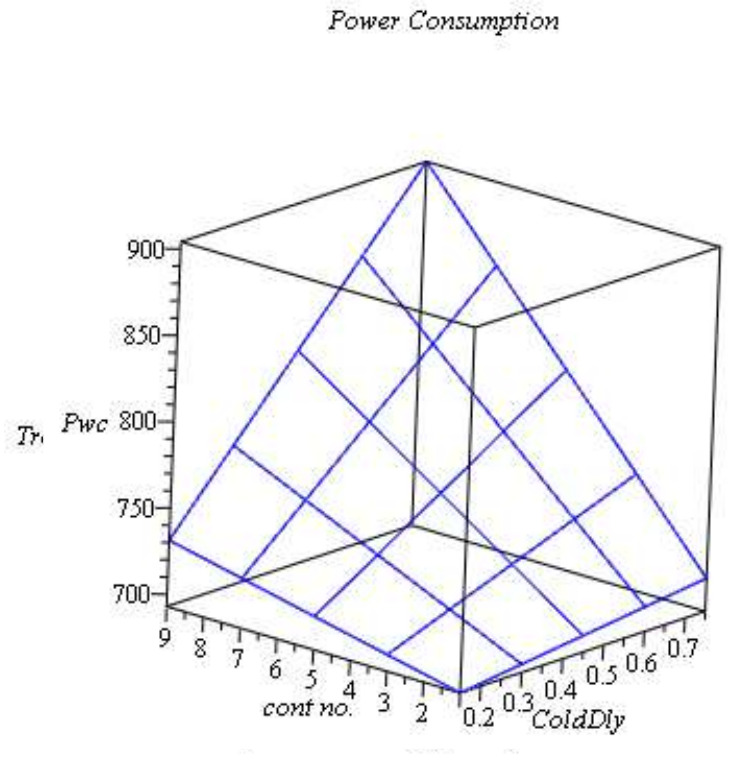}
         \caption{}
         \label{fig8b}
     \end{subfigure}
     \begin{subfigure}[b]{0.3\textwidth}
         \includegraphics[width=5.4cm]{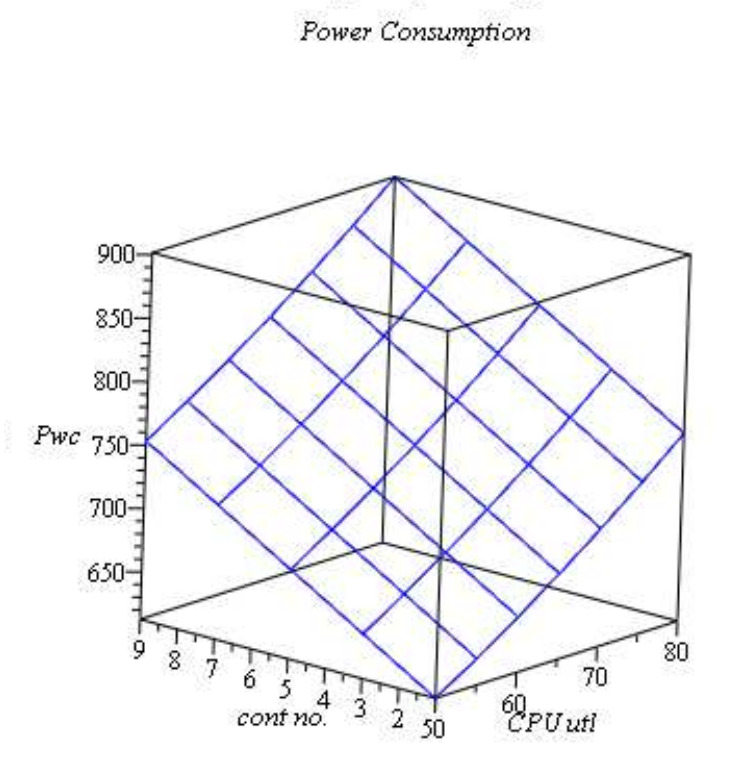}
         \caption{}
         \label{fig8c}
     \end{subfigure}
     \caption{Performance models results (a) Mean response time with varying the number of replicas from 2 to 10. (b) Total power consumption with varying the cold start delay. (c) Total power consumption with varying the number of containers from 1 to 10.} \label{fig8}
\end{figure*}

In the first scenario, we obtained the mean response time of the network with different arrival rates and varying numbers of replicas from $n=1$ to $n=10$ for a function. Figure \ref{fig8} illustrates the effect of event arrival rates on the performance measure of the model. As expected, the mean response time of the network exhibits a growth pattern with increasing arrival rates of events, as shown in Figure \ref{fig8a}. Furthermore, the introduction of additional instances (i.e., replicas of a function) to the network results in a reduction of this response time. This experiment demonstrates the impact of the number of replicas on the delay, which is in line with what we observed in our experimental results. However, it is important to note that increasing the number of replicas also leads to an increase in power consumption.

Therefore, in the second scenario, we explore the power consumption in the network with different cold container startup delays and varying numbers of containers. The analysis of the graph in Figure \ref{fig8b} shows that increasing the number of running containers results in an increase in power consumption due to the cold startup. Additionally, as the cold start delay extends, power consumption also rises. This relationship becomes more apparent when the number of containers is increased. In the next scenario, we fixed the cold start delay while altering the CPU utilisation to investigate how changes in PM utilisation impact power consumption, see Figure \ref{fig8c}. Our observations confirmed that as the server load increases, power consumption rises accordingly. However, when the number of active containers is high, this increase is more pronounced due to the additional power consumed during cold start. 

\section{Modular SDN Prototype Implementation}

In this section, we introduce the software components and technologies employed in the implementation of the serverless SDN/NFV architecture. Our implementation incorporates the integration of three widely recognised open-source projects: 

1) \textit{Open Network Operating System (ONOS)} \cite{Ons2014}, an SDN controller whose architectural design follows a layered structure, including the application layer, core layer, and providers/protocols layer;

2) \textit{general-purpose Remote Procedure Call (gRPC)} \cite{Grp2016}, an RPC system that utilises a protocol buffer and HTTP/2 as its underlying transport protocol and serves as the connection mechanism between microservices; 

3) \textit{Functions as a Service (OpenFaaS)} \cite{Opn2021}, a framework for building event-driven serverless functions on top of containers (with Docker and Kubernetes) and supporting the different architectural components as shown in Figure \ref{fig3}). 


In ONOS, we utilise the Northbound Interfaces (NBI) in the {\it application layer} to receive updates on network events and states. The {\it core part} of ONOS manages network states, maintains an inventory of connected devices, hosts, and links, and provides an overview of the network topology. Additionally, it handles the installation and management of rules in network devices. In the {\it provider/protocol} layer, the Southbound Interface (SBI) encompasses a set of plugins consisting of a provider interface that integrates protocol-specific libraries and a service interface. SBI therefore facilitates interaction with the network environment using different control and configuration protocols. 

We leverage \textit{Mininet} environment \cite{Min2007} to emulate the underlying network infrastructure, including switches and hosts, as well as generate network packets from the data plane.

In order to inform applications about network changes and externalise event processing in ONOS \cite{Cor2019}, we added an event processing layer to enable the division of control plane services into a set of cooperating microservices, see figure \ref{fig4}. This approach allows core services to deliver fundamental functionalities, while supplementary features can be integrated using microservices within the OpenFaaS platform.

We have developed a gRPC server as an ONOS application, leveraging gRPC features such as security, authentication mechanisms and bidirectional streaming. Additionally, gRPC enables automatic code generation for both the server and client sides. The gRPC server interacts with ONOS services, converts the returned values into protobuf format and sends them to the client after serialisation. On the other hand, a gRPC-connector is developed as an external application that acts as the client stub and establishes a connection between ONOS services and OpenFaaS Functions. The latter are invoked by various types of events, including HTTP, which allows for seamless system integration, and made accessible through HTTP endpoints via the Gateway service.

Upon receiving an event from ONOS, the gRPC-connector initiates a query to the Gateway service in order to retrieve the list of available functions. It then creates a mapping between the event topic and the corresponding functions. This mapping enables the triggering of the appropriate function based on the topic associated with the received event.  

\section{Performance Evaluation}
\subsection{Testbed and Experimental Design}
To evaluate the performance of modular SDN/NFV architecture, a prototype of the proposed architecture has been deployed on three Microsoft Azure Virtual Machines (VM), each equipped with 4 VCPUs and 16GB memory, see Table \ref{t2} for details. The first VM hosts ONOS with a gRPC server installed on it, along with Mininet which emulates a network with a number of OpenFlow-enabled switches and hosts. The second VM runs OpenFaaS as a resource pool and network functions host. The third VM has Docker containers deployed on Kubernetes as a resource pool and performs the same tasks as the OpenFaaS functions. These servers run Ubuntu 20.04 LTS and are connected via a 1Gbps LAN network. 

\begin{table}
\caption{Configuration of servers in the experiments.}
\label{t2}       
\centering
\begin{tabular}{l|l||l|l}
\hline\hline
Property & Value & Property & Value \\
\noalign{}\hline\noalign{}
\quad vCPU & 4 & RAM & 16GB  \\
\quad Network & 1000 Mb/s & Ubuntu & 20.04 LTS\\
\quad Kubernetes & 1.18 & Docker & 19.2 \\
\quad ONOS & 2.7.0 & OpenFaaS &  v0.5\\
\quad  Java & 11 & Python & 3.5\\
\hline\hline
\end{tabular}
\end{table}

\begin{figure}
     \begin{subfigure}[b]{0.5\textwidth}
     \hspace{-0.1cm}
         \includegraphics[width=8.5cm]{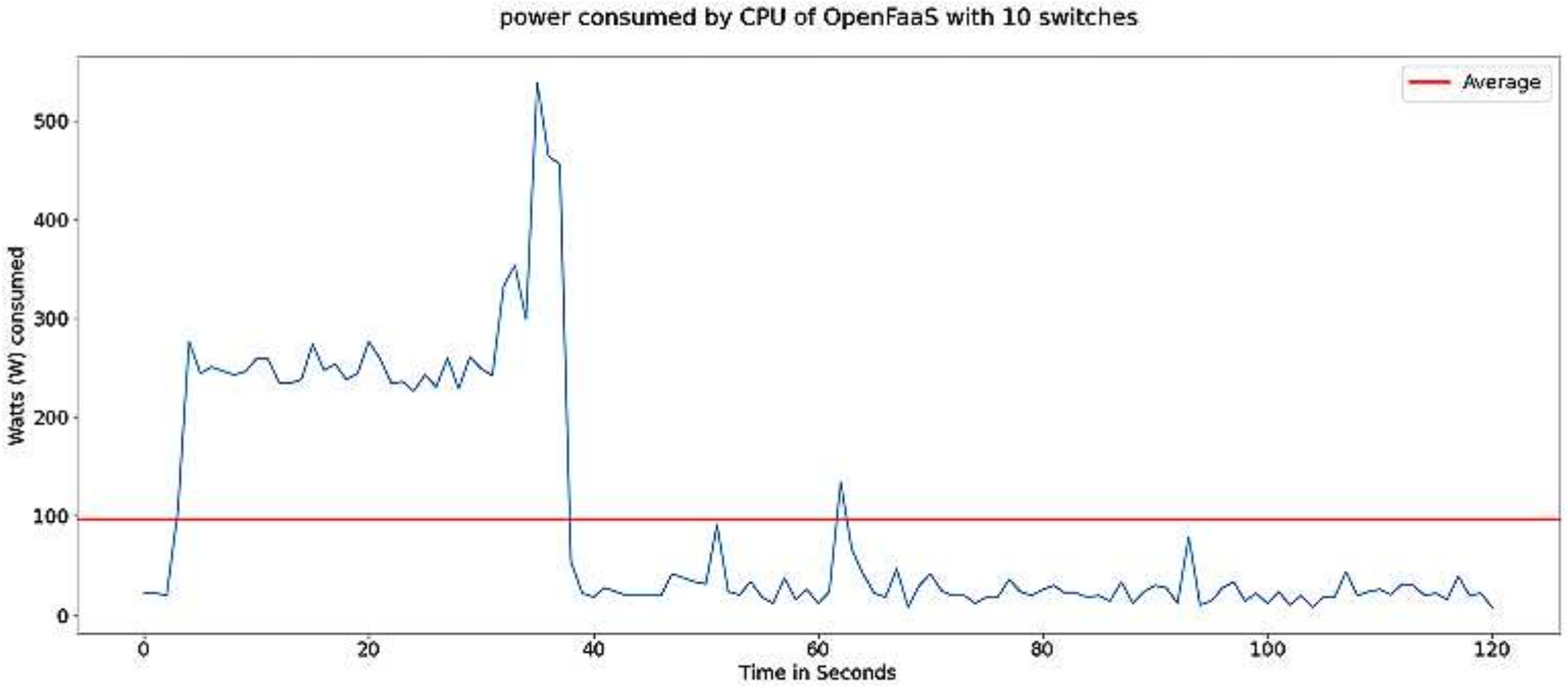}
         \caption{}
         \label{fig5a}
     \end{subfigure}
     \begin{subfigure}[b]{0.5\textwidth}
         \includegraphics[width=8.5cm]{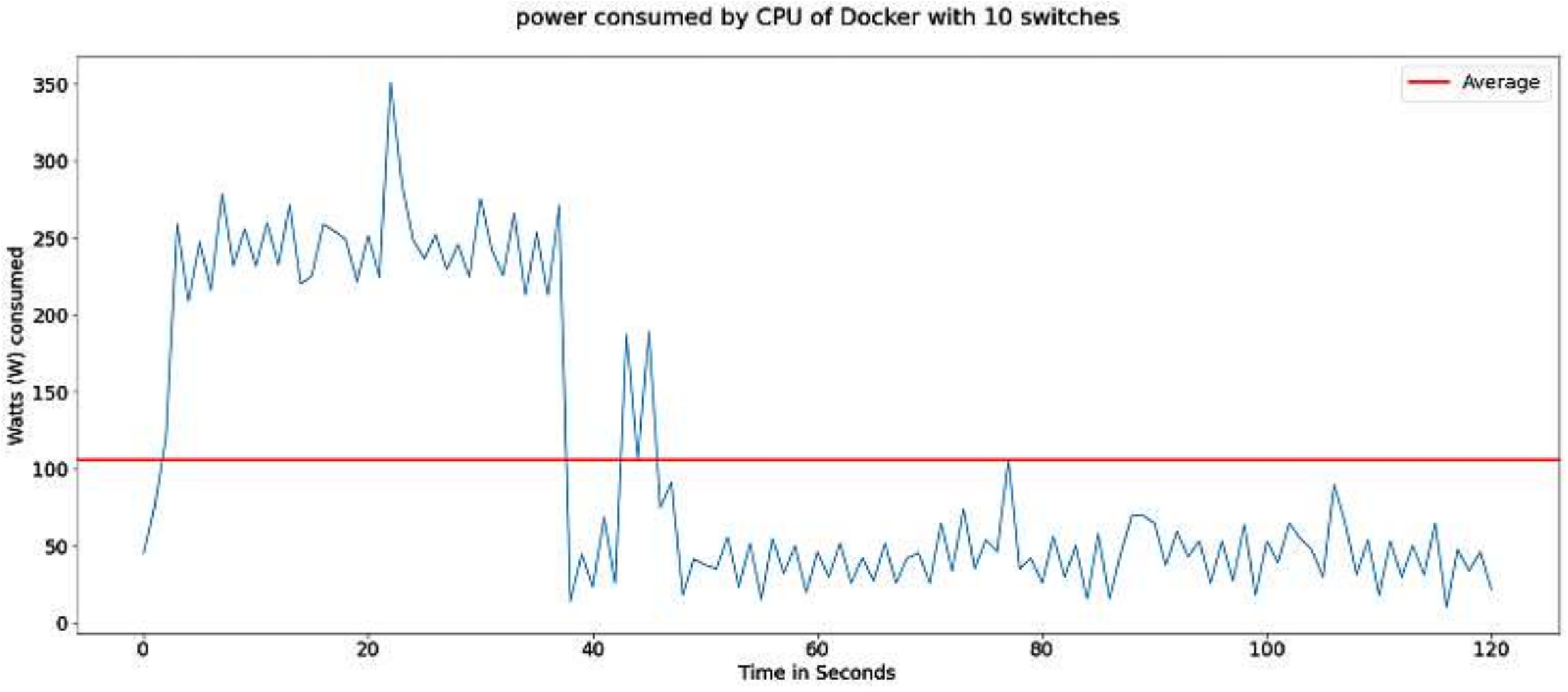}
         \caption{}
         \label{fig5b}
     \end{subfigure}
     \caption{Power consumption of network with 10 switches, (a) OpenFaaS deployment, (b) Docker deployment. }\label{fig5}
\end{figure}

\begin{figure}
     \begin{subfigure}[b]{0.5\textwidth}
         \includegraphics[width=8.5cm]{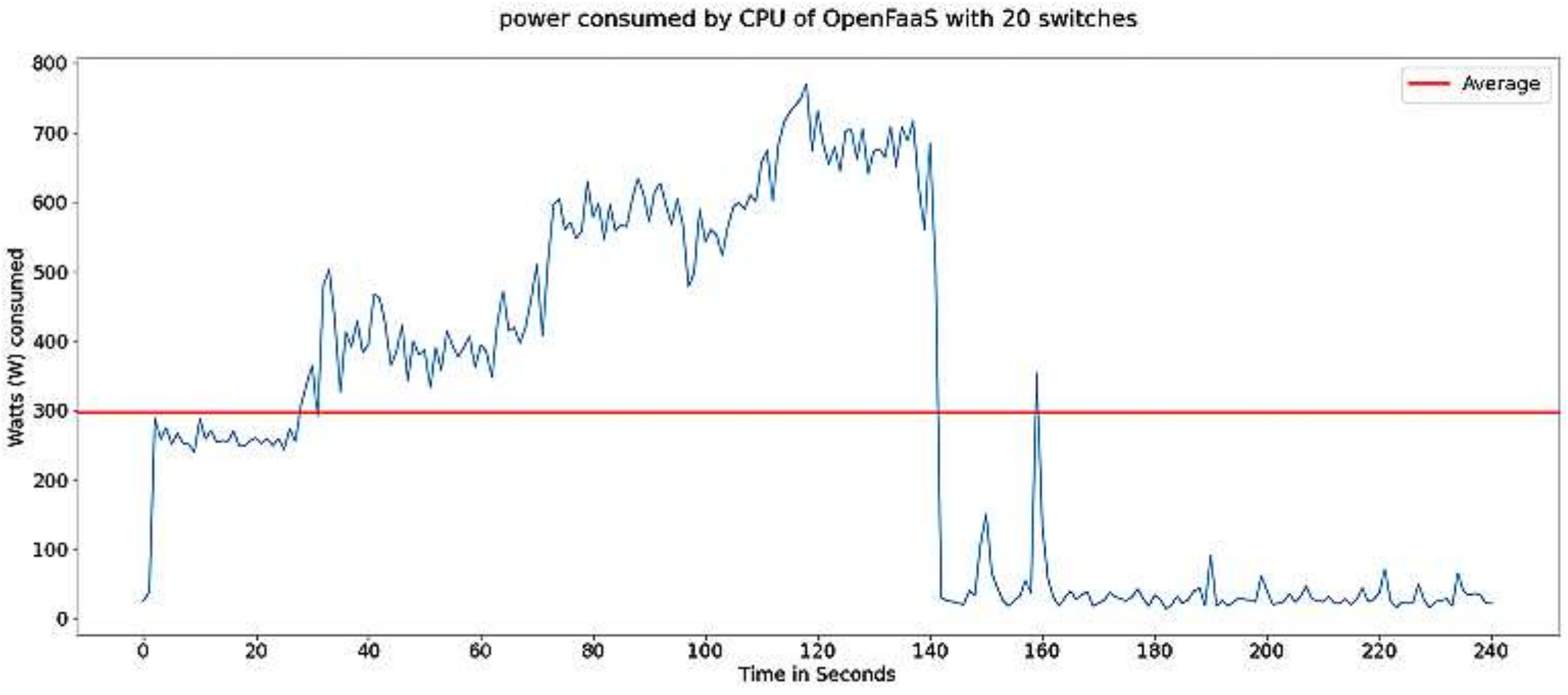}
         \caption{}
         \label{fig6a}
     \end{subfigure}
     \begin{subfigure}[b]{0.5\textwidth}
         \includegraphics[width=8.5cm]{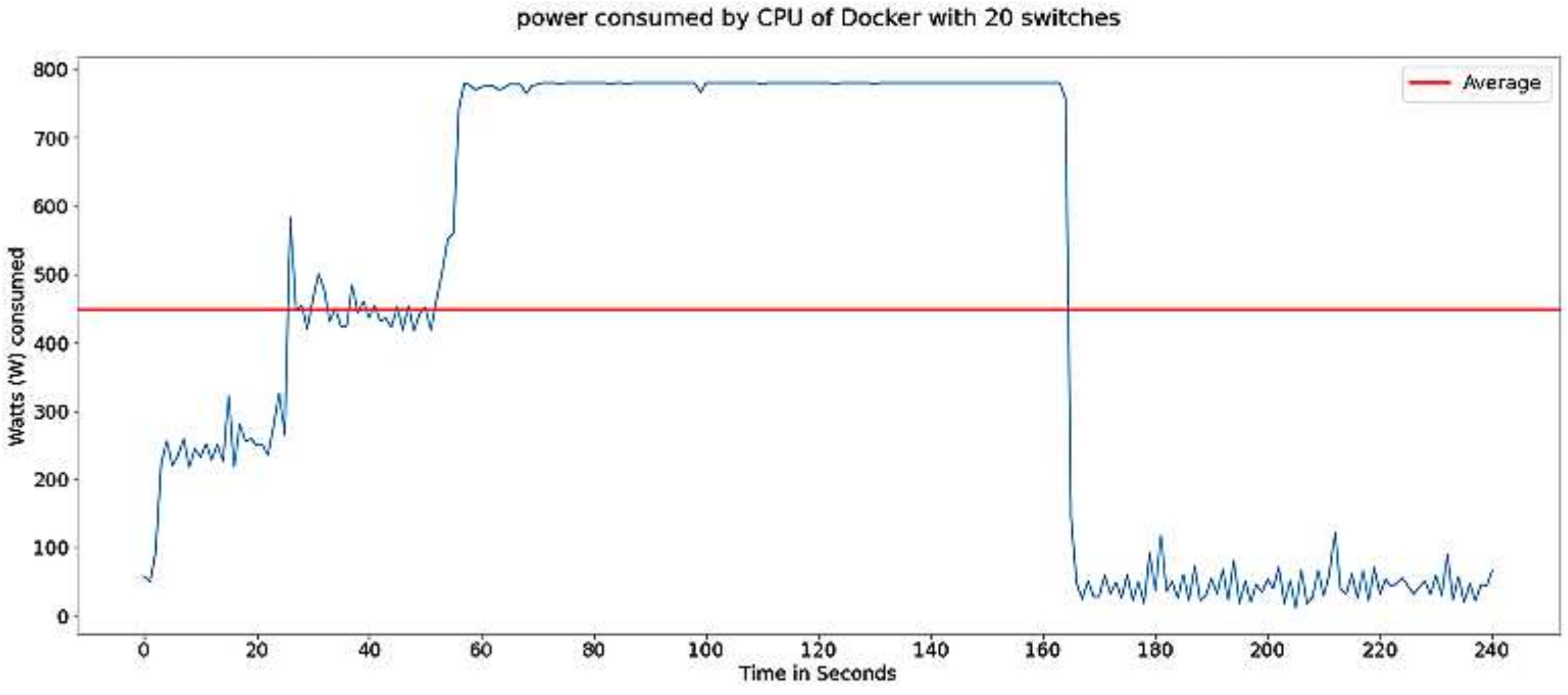}
         \caption{}
         \label{fig6b}
     \end{subfigure}
     \caption{Power consumption of network with 20 switches, (a) OpenFaaS deployment, (b) Docker deployment.}\label{fig6}
\end{figure}


We have conducted the experimental evaluation of the microservice-based SDN architecture through two different scenarios. The objective of the evaluation is to compare the energy consumption and experiment time of the serverless architecture with Docker containers in the context of processing the events coming from ONOS. In these scenarios, once a host sends a packet/event, the switch directs the incoming packet to the SDN controller. The event listening module of the controller serves the packet by sending it to the external application deployed on either a serverless platform or a Docker container via a gRPC channel. ONOS communication with VNFs is handled via gRPC protocol, while OpenFaaS and Kubernetes orchestrate the coming requests by an automatic scaling process. 

To measure power comsumption, the proposed power model in \cite{Fan:2007} is adopted as it is deemed more reliable than, e.g. a tool such as PowerTop \cite{powertop:2023} in a virtualised environment. This model applies the Thermal Design Power (TDP) values of the processors running within the Physical Machines (PM) on Azure, which are Intel Xeon Platinum 8730C, featuring 32 cores and 64 threads with a TDP of 270W. 


\subsection{Experimental Results}

The first scenario involves simulating a network size of 10 virtual OpenFlow-enabled switches and generating a workload using Mininet commands.  The second scenario replicates the same task but with a network size consisting of 20 switches. The commands result in the generation of approximately 10 thousand packets that will be forwarded as requests to the OpenFaaS function or the Docker container, while the set with 10 switches generates around 3 thousand packets. Note that whenever we use OpenFaaS, we are referring to our proposed architecture employed in this experiment.

Figure \ref{fig5} illustrates the power consumption of both platforms, OpenFaaS and Docker, in the context of the first scenario. In this scenario, the server hosting OpenFaaS exhibited an average power consumption of 97 Watts (see figure \ref{fig5a}), whereas the VM running Docker consumed 106 Watts (see figure \ref{fig5b}). These graphs clearly show that the power consumption of OpenFaaS is moderately more efficient compared to Docker containers. However, with an increase in the number of switches to 20, the deviation in performance between OpenFaaS and Docker becomes more noticeable, as depicted in figure \ref{fig6}. During a fixed interval of 120 seconds, the server hosting OpenFaaS exhibited a power consumption of approximately 300 Watts while handling incoming requests (see figure \ref{fig6a}), whereas the server running Docker container consumed about 450 Watts over the same time interval (see figure \ref{fig6b}). This distinction highlights that Docker required approximately 50\% more power consumption than OpenFaaS. Increased power consumption variability in both platforms in high network load is also observed. However, OpenFaaS experience greater stability in its energy usage under higher workloads, which is due to the dynamic allocation and scaling of resources.



\begin{table}[]
    \centering    
\begin{tabular}{c|c|c}
    \textbf{Metrics} &  \textbf{Docker}  & \textbf{OpenFaaS} \\ \hline
    EPC (Watt) & 106 & 97 \\ \hline
    Energy (Joule) & 12,844 & 11,748 \\ \hline
    Load Test Time (sec) & 42 & 37 \\ \hline
    No. of Replicas & 1 & 2 \\ \hline
\end{tabular}    
    \caption{OpenFaaS vs. Docker Comparison - Metrics (10 switches)}
    \label{tab:10switches}
\end{table}

\begin{table}[]
    \centering    
\begin{tabular}{c|c|c}
    \textbf{Metrics} &  \textbf{Docker}  & \textbf{OpenFaaS} \\ \hline
    EPC (Watt) & 448 & 296 \\ \hline
    Energy (Joule) & 108,135 & 71,532 \\ \hline
    Load Test Time (sec) & 168 & 145 \\ \hline
    No. of Replicas & 5 & 5 \\ \hline
\end{tabular}    
    \caption{OpenFaaS vs. Docker Comparison - Metrics (20 switches)}
    \label{tab:20switches}
\end{table}

Tables \ref{tab:10switches} and \ref{tab:20switches} provide a deeper insight into this experiment, presenting details about the Estimated Power Consumption (EPC), energy consumption, experiment duration, and number of replicas. Here, energy consumption is calculated by taking into account power consumption over time. As observed in Table \ref{tab:10switches}, in the first scenario the docker platform creates only one replica of the function, whereas OpenFaaS scales up to two replicas. Consequently, this scaling strategy contributes to a more favorable response time for OpenFaaS (i.e., completing the experiment in 37 seconds in comparison with docker with 42 seconds). However, this variance in scaling does not influence the energy consumption of OpenFaaS, and it is noted that Docker still consumes 9\% more energy than OpenFaaS. The reason is that OpenFaaS is comparatively lighter in terms of creating and managing containers than Docker. In the second scenario, despite both platforms utilising 5 replicas of the function, OpenFaaS completes the experiment earlier than Docker as observed in table \ref{tab:20switches}. Moreover, Docker consumed 51.2\% more energy which indicates a significant improvement in energy consumption of our proposed architecture  compared to the existing platforms.




\section{Conclusion}
In this paper, a microservice-based SDN architecture designed to achieve an automatic and scalable network management approach is introduced. Serverless edge computing is employed to distribute SDN services, making it easy to add new services to the network. An analytical model to approximate the service delivery time and power consumption is presented, as well as an implementation of a prototype of the architecture, accordingly validating its applicability. The evaluation results highlight the capability of this architecture to effectively address energy consumption concerns within SDNs, and indicates a remarkable energy efficiency improvement of nearly 50\% compared to existing platforms (Docker) as well.

With their quantifiable inputs, the analytical model and the prototype implementation are useful in predicting future system behaviour and allow to test the validity of the assumptions in the system and make data-based decisions. The impact of the number of replicas and the power consumption are in line with what was observed in the experimental results. 

One of the objectives of this architecture is to provide fast and cost-efficient services in the context of next-generation industrial networks. The model of the industrial system enables faster time-to-market for new services as it can initiate on-demand virtual networks for providing a set of functionalities on the same hardware platform. Therefore, it suggests a potential to reduce capital and operating costs, especially in the case of short-lived services. By integrating cutting-edge technologies and dynamically adapting to evolving demands, it contributes to building more sustainable and energy-efficient network environments.

Future work includes an investigation of an energy-efficient pluggable solution to scheduling SDN tasks in the serverless computing environment. This can be achieved through the integration of a learning-based models into, e.g. OpenFaaS scheduler \cite{Chiorescu:2023}. As machine learning has been gaining much popularity in recent years, the approach will combine resource management with machine learning algorithms to optimise resource configurations and energy consumption. 

\section*{Acknowledgment}
The authors would like to thank the European Next Generation Internet Program for Open INTErnet Renovation (NGI-Pointer 2) for supporting this work under contract 871528 (EDGENESS Project).

\bibliography{mybiblio.bib} 

\begin{thebibliography}{10}
\providecommand{\url}[1]{#1}
\csname url@samestyle\endcsname
\providecommand{\newblock}{\relax}
\providecommand{\bibinfo}[2]{#2}
\providecommand{\BIBentrySTDinterwordspacing}{\spaceskip=0pt\relax}
\providecommand{\BIBentryALTinterwordstretchfactor}{4}
\providecommand{\BIBentryALTinterwordspacing}{\spaceskip=\fontdimen2\font plus
\BIBentryALTinterwordstretchfactor\fontdimen3\font minus \fontdimen4\font\relax}
\providecommand{\BIBforeignlanguage}[2]{{%
\expandafter\ifx\csname l@#1\endcsname\relax
\typeout{** WARNING: IEEEtran.bst: No hyphenation pattern has been}%
\typeout{** loaded for the language `#1'. Using the pattern for}%
\typeout{** the default language instead.}%
\else
\language=\csname l@#1\endcsname
\fi
#2}}
\providecommand{\BIBdecl}{\relax}
\BIBdecl

\bibitem{Makh2022}
M.~AL-Makhlafi, H.~Gu, A.~Almuaalemi, E.~Almekhlafi, and M.~M. Adam, ``Ribsnet: A scalable high-performance, and cost-effective two-layer based cloud data center network architecture,'' \emph{IEEE Transactions on Network and Service Management}, vol.~20, no.~2, pp. 1676--1690, 2023.

\bibitem{Ban2018}
F.~Bannour, S.~Souihi, and A.~Mellouk, ``Distributed sdn control: Survey, taxonomy and challenges,'' \emph{IEEE Communications Surveys and Tutorials}, vol.~20, no.~1, pp. 333--354, 2018.

\bibitem{Bal2021}
H.~Balakrishnan and et.al, ``Revitalizing the public internet by making it extensible,'' \emph{ACM SIGCOMM Computer Communication Review}, vol.~51, p.~2, 2021.

\bibitem{McC2019}
J.~McCauley, Y.~Harchol, A.~Panda, B.~Raghavan, and S.~Shenker, ``Enabling a permanent revolution in internet architecture,'' in \emph{Proceedings of the ACM Special Interest Group on Data Communication}, ser. SIGCOMM'19.\hskip 1em plus 0.5em minus 0.4em\relax Beijing, China: ACM, 2019, p. 1–14.

\bibitem{Bek2018}
C.~Bektas, S.~Monhof, F.~Kurtz, and C.~Wietfeld, ``Towards 5g: An empirical evaluation of software-defined end-to-end network slicing,'' in \emph{2018 IEEE Globecom Workshops (GC Workshops)}, 2018, pp. 1--6.

\bibitem{Fal2016}
M.~Falkner, A.~Leivadeas, I.~Lambadaris, and G.~Kesidis, ``Performance analysis of virtualized network functions on virtualized systems architectures,'' in \emph{Proc. of the 21st IEEE International Workshop on Computer Aided Modelling and Design of Communication Links and Networks (CAMAD)}.\hskip 1em plus 0.5em minus 0.4em\relax IEEE, Dec. 2016, pp. 71--76.

\bibitem{Ust2020}
R.~F. Ustok, U.~Acar, S.~Keskin, D.~Breitgand, A.~Weit, P.~Drakoulis, A.~Doumanoglou, N.~Zioulis, D.~Zarpalas, P.~Daras, F.~Iadanza, F.~Moscatelli, and G.~Bernini, ``Service development kit for media-type virtualized network services in 5g networks,'' \emph{IEEE Communications Magazine}, vol.~58, no.~7, pp. 51--57, 2020.

\bibitem{Ons2014}
P.~Berde, M.~Gerola, J.~Hart, Y.~Higuchi, M.~Kobayashi, T.~Koide, B.~Lantz, B.~O'Connor, P.~Radoslavov, W.~Snow, and G.~Parulkar, ``Onos: towards an open, distributed sdn os,'' in \emph{Proceedings of the Third Workshop on Hot Topics in Software Defined Networking}, ser. HotSDN'14.\hskip 1em plus 0.5em minus 0.4em\relax Chicago, USA: ACM, 2014, p. 1–6.

\bibitem{Med2014}
J.~Medved, R.~Varga, A.~Tkacik, and K.~Gray, ``Opendaylight: Towards a model-driven sdn controller architecture,'' in \emph{Proceedings of IEEE International Symposium on a World of Wireless}.\hskip 1em plus 0.5em minus 0.4em\relax Sydney, Australia: IEEE, June 2014, pp. 1--6.

\bibitem{Cor2019}
D.~Cormer and A.~Rastegarnia, ``Toward dissagregating the sdn control plane,'' \emph{IEEE Communication Magazine}, vol.~57, no.~10, pp. 70--75, 2019.

\bibitem{Li:2023}
Y.~Li, Y.~Lin, Y.~Wang, K.~Ye, and C.~Xu, ``Serverless computing: State-of-the-art, challenges and opportunities,'' \emph{IEEE Transactions on Services Computing}, vol.~16, no.~2, pp. 1522--1539, 2023.

\bibitem{Adi2019}
P.~Aditya, I.~E. Akkus, A.~Beck, R.~Chen, V.~Hilt, I.~Rimac, K.~Satzke, and M.~Stein, ``Will serverless computing revolutionize nfv?'' \emph{Proceeding of the IEEE}, vol. 107, no.~4, pp. 667--678, 2019.

\bibitem{Djemame:2021}
K.~Djemame, ``Energy efficiency in edge environments: a serverless computing approach,'' in \emph{Economics of Grids, Clouds, Systems, and Services}, K.~Tserpes, J.~Altmann, J.~{\'A}. Ba{\~{n}}ares, O.~Agmon Ben-Yehuda, K.~Djemame, V.~Stankovski, and B.~Tuffin, Eds.\hskip 1em plus 0.5em minus 0.4em\relax Cham: Springer, 2021, pp. 181--184, lecture Notes in Computer Science 13072.

\bibitem{Banaie:2022}
F.~Banaie~Heravan and K.~Djemame, ``A serverless computing platform for software defined networks,'' in \emph{Economics of Grids, Clouds, Systems, and Services}, K.~Tserpes, J.~Altmann, J.~Ba{\~{n}}ares, O.~Agmon Ben-Yehuda, K.~Djemame, and V.~Stankovski, Eds.\hskip 1em plus 0.5em minus 0.4em\relax Izola, Slovenia: Springer, Sep 2022, lecture Notes in Computer Science 13430.

\bibitem{docker2}
D.~Merkel, ``Docker: Lightweight {Linux} containers for consistent development and deployment,'' \emph{Linux Journal}, vol. 2014, no. 239, March 2014.

\bibitem{Opn2021}
``Openfaas - serverless functions, made simple,'' 2023, https://openfaas.com/.

\bibitem{Grp2016}
\BIBentryALTinterwordspacing
grpc. (2023) A high-performance, open-source, general-purpose rpc framework.[online]. available:. [Online]. Available: \url{https://github.com/grpc}
\BIBentrySTDinterwordspacing

\bibitem{rltd1}
S.~Rout, K.~S. Sahoo, S.~S. Patra, B.~Sahoo, and D.~Puthal, ``Energy efficiency in software defined networking: A survey,'' \emph{SN Computer Science}, vol.~2, p.~4, 2021.

\bibitem{Maity:2023}
I.~Maity, R.~Dhiman, and S.~Misra, ``Enplace: Energy-aware network partitioning for controller placement in sdn,'' \emph{IEEE Transactions on Green Communications and Networking}, vol.~7, no.~1, pp. 183--193, 2023.

\bibitem{Oliveira:2021}
T.~F. Oliveira, S.~Xavier-de Souza, and L.~F. Silveira, ``Improving energy efficiency on sdn control-plane using multi-core controllers,'' \emph{Energies}, vol.~14, no.~11, 2021.

\bibitem{Priyadarsini:2020}
M.~Priyadarsini, S.~Kumar, and P.~Bera, ``An energy-efficient load distribution framework for sdn controllers,'' \emph{Computing}, vol. 102, p. 2073–2098, 2020.

\bibitem{Alhindi:2022}
A.~Alhindi, K.~Djemame, and F.~Banaie, ``On the power consumption of serverless functions: an evaluation of openfaas,'' in \emph{Proceedings of the 15th IEEE/ACM International Conference on Utility and Cloud Computing (UCC 2022), Vancouver, IEEE}, 2022.

\bibitem{kube2}
\BIBentryALTinterwordspacing
{The Kubernetes Authors}. (2023) Kubernetes documentation. [Online]. Available: \url{https://kubernetes.io/docs/home/}
\BIBentrySTDinterwordspacing

\bibitem{Cicconetti:2021}
C.~Cicconetti, M.~Conti, and A.~Passarella, ``A decentralized framework for serverless edge computing in the internet of things,'' \emph{IEEE Transactions on Network and Service Management}, vol.~18, no.~2, pp. 2166--2180, 2021.

\bibitem{Shen:2020}
J.~Shen, H.~Yu, Z.~Zheng, C.~Sun, M.~Xu, and J.~Wang, ``Serpens: A high-performance serverless platform for nfv,'' in \emph{2020 IEEE/ACM 28th International Symposium on Quality of Service (IWQoS)}, 2020, pp. 1--10.

\bibitem{Tzenetopoulos:2021}
A.~Tzenetopoulos, C.~Marantos, G.~Gavrielides, S.~Xydis, and D.~Soudris, ``Fade: Faas-inspired application decomposition and energy-aware function placement on the edge,'' in \emph{Proceedings of the 24th International Workshop on Software and Compilers for Embedded Systems}, ser. SCOPES'21.\hskip 1em plus 0.5em minus 0.4em\relax Eindhoven, Netherlands: ACM, 2021, p. 7–10.

\bibitem{Jia:2021}
X.~Jia and L.~Zhao, ``Raef: Energy-efficient resource allocation through energy fungibility in serverless,'' in \emph{Proceedings of the 27th IEEE International Conference on Parallel and Distributed Systems (ICPADS)}.\hskip 1em plus 0.5em minus 0.4em\relax Beijing, China: IEEE, Dec 2021, pp. 434--441.

\bibitem{Zhang:2022}
L.~Zhang, Y.~Pu, C.~Xu, D.~Liu, Z.~Lin, X.~Hou, P.~Yang, S.~Yue, C.~Li, and M.~Guo, ``Cloud-native server consolidation forÂ energy-efficient faas deployment,'' in \emph{Network and Parallel Computing}, S.~Liu and X.~Wei, Eds.\hskip 1em plus 0.5em minus 0.4em\relax Cham: Springer Nature, 2022, pp. 120--126.

\bibitem{Das:2020}
A.~Das, A.~Leaf, C.~A. Varela, and S.~Patterson, ``Skedulix: Hybrid cloud scheduling for cost-efficient execution of serverless applications,'' in \emph{Proceedings of the 13th IEEE International Conference on Cloud Computing (CLOUD)}.\hskip 1em plus 0.5em minus 0.4em\relax IEEE, 2020, pp. 609--618.

\bibitem{Fan:2020}
\BIBentryALTinterwordspacing
D.~Fan and D.~He, ``A scheduler for serverless framework base on kubernetes,'' in \emph{Proceedings of the 2020 4th High Performance Computing and Cluster Technologies Conference \& 2020 3rd International Conference on Big Data and Artificial Intelligence}, ser. HPCCT \& BDAI '20.\hskip 1em plus 0.5em minus 0.4em\relax Qingdao, China: ACM, 2020, p. 229–232. [Online]. Available: \url{https://doi.org/10.1145/3409501.3409503}
\BIBentrySTDinterwordspacing

\bibitem{Arz2022}
S.~Arzo, D.~Scotece, R.~Bassoli, D.~Barattini, F.~Granelli, L.~Foschini, and F.~Fitzek, ``Msn: A playground framework for design and evaluation of microservices-based sdn controller,'' \emph{Journal of Network and Systems Management}, vol.~30, 2022.

\bibitem{Tan2020}
M.~Tang and V.~W.~S. Wong, ``Deep reinforcement learning for task offloading in mobile edge computing systems,'' \emph{IEEE Transactions on Mobile Computing}, vol.~21, pp. 1985--1997, 2020.

\bibitem{Tan2022}
Q.~Tang, R.~Xie, F.~R. Yu, T.~Chen, R.~Zhang, T.~Huang, and Y.~Liu, ``Distributed task scheduling in serverless edge computing networks for the internet of things: A learning approach,'' \emph{IEEE Internet of Things Journal}, pp. 1--1, April 2022.

\bibitem{Gri2010}
G.~Grimmett and D.~Stirzaker, \emph{Probability and Random Processes}, 3rd~ed.\hskip 1em plus 0.5em minus 0.4em\relax Oxford: Univ. Press, 2010.

\bibitem{Rio2014}
O.~Rioul and J.~Magossi, ``On shannon’s formula and hartley’s rule: Beyond the mathematical coincidence,'' \emph{Entropy}, vol.~16, no.~9, p. 4892–4910, 2014.

\bibitem{Ros2014}
M.~Ross, \emph{Introduction to Probability Models}, 13th~ed.\hskip 1em plus 0.5em minus 0.4em\relax Academic Press, June 2023.

\bibitem{mohan:2019}
A.~Mohan, H.~Sane, K.~Doshi, S.~Edupuganti, N.~Nayak, and V.~Sukhomlinov, ``Agile cold starts for scalable serverless,'' in \emph{11th {USENIX} Workshop on Hot Topics in Cloud Computing (HotCloud 19)}.\hskip 1em plus 0.5em minus 0.4em\relax Renton, WA: {USENIX} Association, Jul. 2019.

\bibitem{Zho2022}
Z.~Zhou, M.~Shojafar, and M.~Alazab, ``Iecl: An intelligent energy consumption model for cloud manufacturing,'' \emph{IEEE Transactions on Industrial Informatics}, vol.~18, no.~12, pp. 8967--8976, 2022.

\bibitem{Zha2018}
H.~Zhao, F.~L. J.~Wang, Q.~Wang, W.~Zhang, and Q.~Zheng, ``Power-aware and performance-guaranteed virtual machine placement in the cloud,'' \emph{IEEE Transactions on Parallel and Distributed Systems}, vol.~29, p.~6, 2018.

\bibitem{Aldossary:2019}
M.~Aldossary, K.~Djemame, I.~Alzamil, A.~Kostopoulos, A.~Dimakis, and E.~Agiatzidou, ``Energy-aware cost prediction and pricing of virtual machines in cloud computing environments,'' \emph{Future Generation Computer Systems}, vol.~93, pp. 442--459, 2019.

\bibitem{Map2015}
\BIBentryALTinterwordspacing
``Maplesoft,'' 2023. [Online]. Available: \url{http://www.maplesoft.com/products/maple}
\BIBentrySTDinterwordspacing

\bibitem{Ste2022}
M.~S. Markus and et. al., ``Tppfaas: Modeling serverless functions invocations via temporal point processes,'' \emph{IEEE Access}, vol.~10, pp. 9059--9084, 2022.

\bibitem{Mah2022}
N.~Mahmoudi and H.~Khazaei, ``Performance modeling of serverless computing platforms,'' \emph{IEEE Transactions on Cloud Computing}, vol.~10, p.~4, 2022.

\bibitem{Sri2018}
A.~Sriraman and T.~F. Wenisch, ``Mu suite: A benchmark suite for microservices,'' \emph{in Proc. IEEE Int. Symp. Workload Characterization}, vol. 1–12, 2018.

\bibitem{Min2007}
\BIBentryALTinterwordspacing
``Mininet,'' 2023. [Online]. Available: \url{http://mininet.org/}
\BIBentrySTDinterwordspacing

\bibitem{Fan:2007}
X.~Fan, W.~Weber, and L.~Barroso, ``Power provisioning for a warehouse-sized computer,'' in \emph{Proc. of the 34th Annual International Symposium on Computer Architecture}.\hskip 1em plus 0.5em minus 0.4em\relax NY, USA: ACM, 2007, p. 13–23.

\bibitem{powertop:2023}
``Powertop,'' 2023, https://wiki.archlinux.org/title/powertop.

\bibitem{Chiorescu:2023}
R.~Chiorescu and K.~Djemame, ``Scheduling energy-aware multi-function serverless workloads in openfaas,'' in \emph{Proc. of the 16th IEEE/ACM International Conference on Utility and Cloud Computing (UCC 2023)}.\hskip 1em plus 0.5em minus 0.4em\relax Taormina (Messina), Italy: IEEE/ACM, Dec. 2023.

\end{thebibliography}
\bibliographystyle{IEEEtran}

\vspace{-2cm}

\begin{IEEEbiography}[{\includegraphics[width=0.9in,height=1.25in,clip,keepaspectratio]{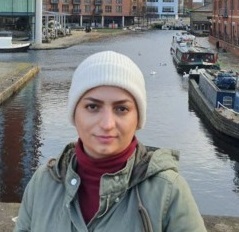}}]{Fatemeh Banaie} is a Research Fellow in Computer Science at the University of Huddersfield. She completed her MSc and Ph.D. in Computer Engineering at the Ferdowsi University of Mashhad (FUM). She previously held a Research Fellow position at the University of Leeds (2021-2023). Her research interests are the Internet of Things, Performance Evaluation, and Edge computing.
\end{IEEEbiography}

\vspace{-1.3cm}
\begin{IEEEbiography}
[{\includegraphics[width=0.9in,height=1.25in,clip,keepaspectratio]{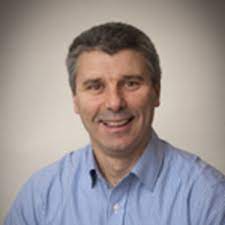}}]{Karim Djemame} is a Professor of Distributed Systems at the University of Leeds. He received a PhD from Glasgow University in 1999. His research interests are in cloud/edge computing, energy-efficient computing systems and performance evaluation. He is a member of IEEE.

\end{IEEEbiography}
\vspace{-1.3cm}

\begin{IEEEbiography}
[{\includegraphics[width=0.9in,height=1.25in,clip,keepaspectratio]{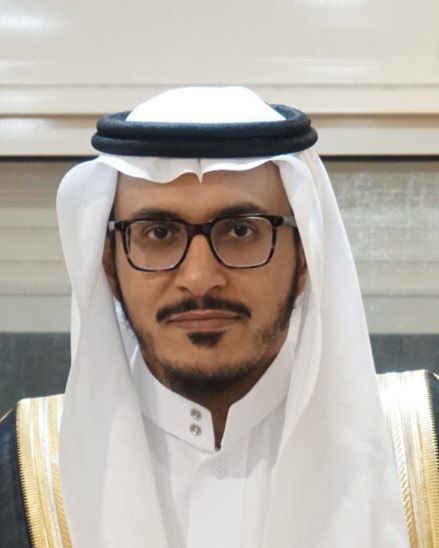}}]{Abdulaziz Alhindi} is working toward the PhD degree in Computer Science at the University of Leeds. He completed his MSc in Computer Science at the University of Leeds in 2022. He has been a lecturer at the Computer Science Department at Qassim University, Kingdom of Saudi Arabia since 2011. His research interests include Serverless computing, SDN, and energy efficiency.
\end{IEEEbiography}

\vspace{-1.3cm}

\begin{IEEEbiography}
[{\includegraphics[width=0.9in,height=1.25in,clip,keepaspectratio]{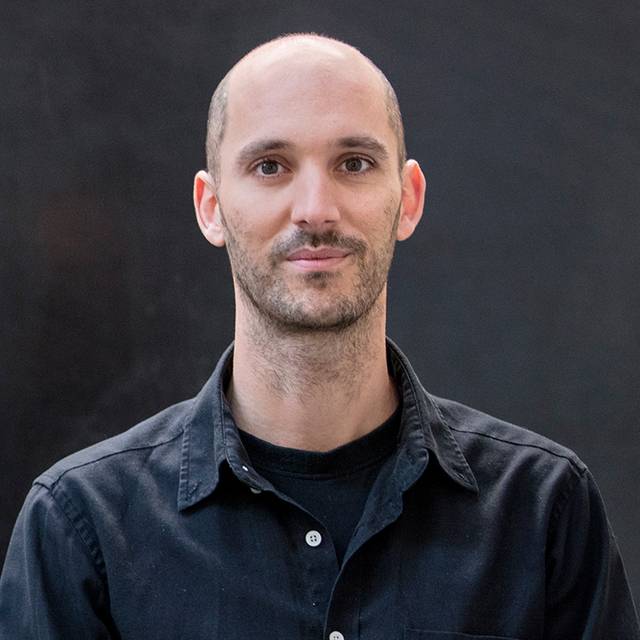}}]{Vasilios Kelefouras} is a Lecturer (Assistant Professor) in computer science at the University of Plymouth (UK). His main research interests are in the areas of compiler optimisations for High-Performance Computing (HPC), loop transformations, data movement optimisation in cache memories, optimisation of Matrix/Tensor computations, data compression methods for Deep Neural Networks and task scheduling for HPC. 
\end{IEEEbiography}

\vfill

\end{document}